\documentclass[sigconf]{acmart}

\usepackage{booktabs}
\usepackage{color}
\usepackage{amsmath}
\usepackage{dirtytalk}

\newcommand{\note}[1]{{\color{red} {\bf #1}}}

\copyrightyear{2019}
\acmYear{2019} 
\setcopyright{iw3c2w3}
\acmConference[WWW '19 Companion]{Companion Proceedings of the 2019 World Wide Web Conference}{May 13--17, 2019}{San Francisco, CA, USA}
\acmBooktitle{Companion Proceedings of the 2019 World Wide Web Conference (WWW '19 Companion), May 13--17, 2019, San Francisco, CA, USA}
\acmPrice{}
\acmDOI{10.1145/3308560.3316502}
\acmISBN{978-1-4503-6675-5/19/05}

\usepackage{balance}

\begin{document}

%
% The ''title'' command has an optional parameter, allowing the author to define a ''short title'' to be used in page headers.
\title{Characterizing Activity on the Deep and Dark Web}

%
% The ''author'' command and its associated commands are used to define the authors and their affiliations.
% Of note is the shared affiliation of the first two authors, and the ''authornote'' and ''authornotemark'' commands
% used to denote shared contribution to the research.
\author{Nazgol Tavabi$^{1}$,Nathan Bartley$^{1}$,Andr\'es Abeliuk$^{1}$,Sandeep Soni$^{2}$,Emilio Ferrara$^{1}$,Kristina Lerman$^{1}$}
%\authornote{Both authors contributed equally to this %research.}
%\email{webmaster@marysville-ohio.com}
\affiliation{%
\institution{1. University of Southern California, Information Sciences Institute}
\institution{2. Georgia Institute of Technology}
% \streetaddress{4676 Admiralty Way}
% \city{Marina Del Rey}
% \state{CA (USA)}
% \postcode{90292}
}

\renewcommand{\shortauthors}{N. Tavabi et al.}

%
% By default, the full list of authors will be used in the page headers. Often, this list is too long, and will overlap
% other information printed in the page headers. This command allows the author to define a more concise list
% of authors' names for this purpose.
%\renewcommand{\shortauthors}{}

%
% The abstract is a short summary of the work to be presented in the article.
\begin{abstract}
The deep and darkweb (d2web) refers to limited access web sites that require registration, authentication, or more complex encryption protocols 
%either registration and authentication, or more complex encryption protocols, 
to access them. These web sites serve as hubs for a variety of illicit activities: to trade drugs, stolen user credentials, hacking tools, and to coordinate attacks and manipulation campaigns. Despite its importance to cyber crime, the d2web has not been systematically investigated. In this paper, we study a large corpus of messages posted to 80 d2web forums over a period of more than a year. We identify topics of discussion using LDA and use a non-parametric HMM to model the evolution of topics across forums. Then, we examine the dynamic patterns of discussion and identify forums with similar patterns. We show that our approach surfaces hidden similarities across different forums and can help identify anomalous events in this rich, heterogeneous data. 
\end{abstract}

\ccsdesc[500]{Mathematics of computing~Time series analysis}
\ccsdesc[500]{Mathematics of computing~Exploratory data analysis}
\ccsdesc[300]{Mathematics of computing~Bayesian nonparametric models}
\ccsdesc[300]{Mathematics of computing~Cluster analysis}
\ccsdesc[500]{Information systems~Deep web}
\ccsdesc[300]{Information systems~Clustering}
\ccsdesc[300]{Information systems~Web crawling}
\ccsdesc[300]{Security and privacy~Social aspects of security and privacy}

%\ccsdesc[500]{Computer systems organization~Embedded systems}

%
% Keywords. The author(s) should pick words that accurately describe the work being
% presented. Separate the keywords with commas.
\keywords{Darkweb; Deepweb; D2web; Cyber Crime; Cyber Security; LDA; Non-Parametric HMM; Beta Process; multivariate Time Series; Cluster Time Series}

%
% A ''teaser'' image appears between the author and affiliation information and the body 
% of the document, and typically spans the page. 

%
% This command processes the author and affiliation and title information and builds
% the first part of the formatted document.
\maketitle

\section{Introduction}

The web that most people are familiar with---the open and searchable internet  of social media sites, online merchants, newspapers and the like---represents just a tiny fraction of the internet. Much larger portions of internet data remain buried within the ``deepweb''~\cite{hurlburt2017shining}, a term that refers to private corporate intranets and databases, dynamically-generated web pages, and limited access content, such as online academic journals. While %some of this content is open,
only some of the content is hidden behind encrypted protocols in .onion domains, access to the remaining ``open'' content is generally restricted, requiring registration and authentication~\cite{shakarian2016exploring}.  Some portion of the deepweb, also known as the ``darkweb,'' serves as a hub for all kinds of illicit activities. Malicious actors congregate virtually on dark web forums and marketplaces to trade illicit information, goods and services, including ransomware, exploits, hacking tools, stolen media, user credentials, fake ids, prescription medicines and illegal drugs. In addition to serving as a marketplace for these goods, the deep and dark web provides a venue for malicious actors to coordinate cyber attacks~\cite{robertson2017darkweb}  and terrorist activity~\cite{chen2008uncovering}. 
The growing popularity of the marketplaces within the deepweb can be attributed to the elimination of the risk of violence since there is limited, if any, physical interaction between the buyers and the sellers. Another reason is the use of encrypted protocols to preserve anonymity, encouraging people to express themselves without the risk of getting caught by law enforcement nor being censored by the moderators of a web site~\cite{soska2015measuring}.
%As our lives become more interconnected through the World Wide Web (WWW), the threat of malicious web activities becomes more alarming. Many deep/darkweb (D2web) forums  and clearnet websites have become hosts to malicious discussions about human trafficking, selling drugs and weapons, hacking tools and techniques, terrorist activities and the like \cite{hurlburt2017shining}.

Given the threat posed by these malicious actors, observing their activities on the deep and dark web (d2web) may provide valuable clues both for anticipating and preventing cyber attacks as well as mitigating the fallout from data breaches. However, picking out useful signals in the vast, dynamic and heterogeneous environment of the d2web can be challenging. 

In this paper, we use Latent Dirichlet Allocation (LDA) \cite{blei2003latent} to analyze a large heterogeneous text corpus from the d2web. To understand the dynamics of discussions, %topics, 
%we mapped the LDA outputs into a one dimensional time series representing different dynamic states over the topic variations. Specifically, the approach proposed in this paper models the dynamics of discussions on the d2web  forums using a
we use a 
non-parametric hidden Markov model~\cite{fox2014joint}---the \textit{Beta Process HMM}. In this approach every forum is represented as a multivariate time series, where variables are the topics found by LDA, and is fed into a Beta Process HMM (BP-HMM). This BP-HMM then finds the shared states among forums, where each state is a distribution over topics. This helps track discussions on different forums and identify anomalous behavior or important events. This approach can also be used to find forums relevant to a specific subject, i.e., forums or time periods within forums where users  discuss specific topics, such as hacking techniques and cyber security related issues. This method can also cluster forums into meaningful groups.

We test this framework on data consisting of posts published on 80 D2web forums from 2016 to mid 2017, on topics such as exploits and hacking techniques, selling prescription and non-prescription drugs, and creating fake ids. Overall this paper makes the following contributions: \begin{itemize}
    \item Using LDA, we characterize the content of $80$ d2web forums where illicit activities are discussed.  
    \item We describe an application of a non-parametric HMM model to learn the forum's shared topic dynamics based on the results obtained from LDA.%multivariate LDA topic dimensions.
    \item  We use the learned shared behaviors as a compact representation of these time series to cluster forums into groups with distinct characteristics and analyze learned latent behavioral structures to gain more insight into this data.
    %\item We use cross entropy to further analyze forum dynamics. 
    \item We rank forums based on how their topics of discussions are likely to change.  
  \item Finally, we present case studies revealing how our approach can be used to to identify anomalous activity in d2web data.
\end{itemize}
%   In this approach each forum is modeled as a multivariate time series of LDA topics and Beta Process Autoregressive HMM (BP-AR-HMM) identifies the shared states between forums. 
\iffalse
\note{
\begin{itemize}
    \item Problem description: finding order in dynamic, noisy, heterogeneous data. 
    \item State of the art: LDA, Dynamic LDA, Clustering
    \item What's new: HMMs, dynamic analysis as tool for imposing structure on the dark web
    \item Applications: find 1) similar forums, 2) anomalies (e.g., marketplace shutdowns), 3) compare languages, 4) malicious innovators, 
\end{itemize}
}

\fi
\section{Related Work}
For an overview of the darkweb and deepweb and the challenges these sites pose for researchers and for law enforcement, please see \cite{hurlburt2017shining,gehl2016power}.  
 Researchers have leveraged d2web content in specific applications, but to the best of our knowledge, few have attempted to systematically characterize the topics and dynamics of d2web discussions. %taken the problem of dealing with it's heterogeneity into consideration. 
 For example, Xu et al.~\cite{xu2008topology} analyzed the topology and structure of darkweb networks. Soska et al.~\cite{soska2015measuring} analyzed 16 different marketplaces to extract information on goods being sold and money transactions, then trained classifiers on them.
 Tavabi et al.~\cite{tavabi2018darkembed} and Almukaynizi et al.~\cite{almukaynizi2017proactive} used features from d2web discussions to predict which new vulnerability will be exploited.  %\cite{goyal2018discovering} used occurrence frequency of important cyber security-related keywords %and 
Goyal et al.~\cite{goyal2018discovering} and Deb et al.~\cite{info9110280} used the frequency of important cyber security-related keywords and the sentiment of d2web posts respectively to predict cyber attacks. Along similar lines, \cite{chen2008sentiment} utilized sentiment to analyze communications of extremists on the darkweb and \cite{al2015bisal} developed a bilingual (English and Arabic) sentiment analysis lexicon for cyber security and radicalism on darkweb forums. 

 In this paper we propose a statistical approach to analyze discussions of multiple malicious forums by extracting their similarities and their differences through learning hidden Markov models on topic weights obtained from LDA. Latent Dirichlet Allocation (LDA) proposed by Blei et al.~\cite{blei2003latent} is a powerful ubiquitous tool which allows documents to be explained by latent topics. LDA has shown to be effective in the analysis of the darkweb \cite{l2011topic,rios2012dark}. However, the dynamics of darkweb topics, which can give deeper insight into this data, has not been analyzed previously. Rios et al.~\cite{rios2012dark} used LDA to detect overlapping communities in the darkweb and L'huillier et al.~\cite{l2011topic} applied LDA to extract key members in darkweb forums. Extensions to LDA, like Dynamic Topic Models \cite{blei2006dynamic}, have been designed to model evolution of topics across time. However, such models have strong memory requirements making them impractical for a large heterogeneous corpus like the d2web. An alternative approach is performing LDA on the entire corpus and observing fluctuation of weights across time, similar to approaches set forth by \cite{tirunillai2014mining,linstead2008application}   and many others. Variants of LDA which combine Hidden Markov Models have also been proposed. The approach proposed by Griffiths et al. ~\cite{griffiths2005integrating} models both semantic and syntactic dependencies of documents by defining one state in HMM as the semantic state modeled by LDA and other states as syntactic components. Gruber et al.~\cite{gruber2007hidden} modeled the relationship between words in a document with an HMM. Their model assumes words in the same sentence have the same topic, and successive sentences are more likely to have the same topics, where in the original LDA paper words in a document are assumed to be independent and documents are modeled as bag of words. These variations were proposed to better capture LDA topics, although Hidden Markov models can also be used to identify shifts and variations in the topics discussed, which is the approach of this work. HMM states can help us recognize important events and anomalies. %and find forums relevant to a specific subject by extracting common states.
 %nathan - i dont think this matters in the related work
 %In general this could serve as a tool to analyze a heterogeneous dataset.%or more precisely Markov switching autoregressive models, each state in HMM is modeled as an autoregressive process, identify shifts in the topics discussed which signal events and help analyze the dataset.
 \begin{figure}
%\vspace{14pc}
\includegraphics[width=0.8\columnwidth]{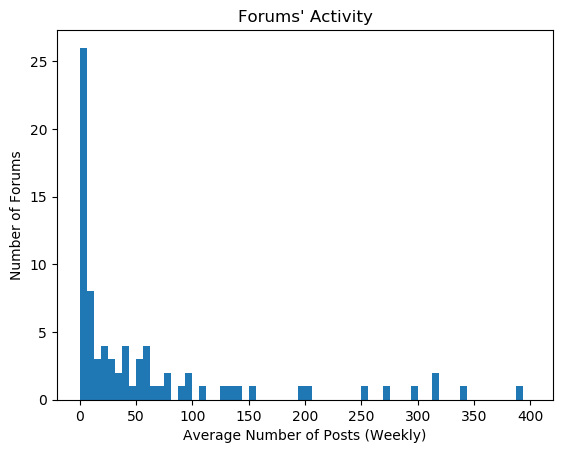}
\caption{Histogram of activity level in different forums.}
\label{fig:activity}
\end{figure}
 \section{Methods}

\subsection{D2Web Data Collection}

%We get the data from Cyr3con, consists of X forums covering different areas including darkweb marketplaces, discussions about pen testing, hardware and software exploits for different platforms (including Playstation, various Linux distros, Windows), and the sharing of confidential information.

%We pull all the posts (ranging in dates from 2016 (likely a datetime issue) to 2018 (when in 2018?) and classify the language using the langid implemented by \cite{lui2012langid}.  
%\note{todo: Nathan}

%\note{nathan: what do we do with this citation}
The d2web data we use in the study was collected using the crawling infrastructure described in %\note{citation issue}\cite{anonimized}.
~\cite{robertson2017darkweb,nunes2016darknet}.
This crawling infrastructure uses anonymization protocols, such as Tor and I2P, to access darkweb sites, and handles authentication to access  non-indexed deepweb sites on the Internet.  The infrastructure includes lightweight crawlers and parsers that are focused on specific sites related to malicious hacking and/or online financial fraud. These sites represent forums and discussion boards where people mostly discuss cyber crime and fraud, although other illicit activities are discussed as well, such as the sale of drugs and other stolen goods. There are also a handful of forums crawled that are on the clearnet but which are mostly white hat (i.e., involved with ethical hacking and/or professional cybersecurity). We include these forums in our analysis primarily to help identify other forums that might discuss similar topics. In all, crawlers scraped data from over 250 d2web forums. The most common languages in which the posts were written were English (accounting for 37.8\% of all posts),  Russian  (22.4\% of all posts), and Chinese (15.4\% of all posts).  Other languages, such as Spanish, Arabic, and Turkish were less frequent, each accounting for less than 7\% of the posts. For this analysis we only focus on English posts, though the same structure could be used for multiple languages. Filtering out the non-English posts brings the number of forums down to 155.
We pre-processed the posts using NLTK \cite{BirdKleinLoper09}, SpaCy \cite{spacy2}, and scikit-learn to remove stopwords, tokenize each post, and filter tokens by post frequency to remove frequent words. This gives us a corpus of 1.33 million posts. % identified as either English, Russian, Spanish, Turkish, or Arabic. In total we analyze approximately 3 million documents. 
%After pre-processing and temporal filtering% out empty documents
%, we analyze approximately 482,000 posts in our d2web corpus (from a larger pool of approximately 1.33 million posts).

\subsection{Modeling Topics of Discussion}
We applied a popular statistical technique known as Latent Dirichlet Allocation (LDA)~\cite{blei2003latent} to learn the topics of the English-language posts. LDA is used to decompose documents into latent topics, where each topic is a distribution over words, intended to capture the semantic content of documents. In this model each document is treated as a bag-of-words. Once we learn the model we can represent documents as distributions over a fixed number of topics. Doing so gives us low-dimensional representations of documents. %and is assumed to have its own distribution over the same set of topics
%We applied a popular statistical technique known as Latent Dirichlet Allocation (LDA)~\cite{blei2003latent}, to learn the topics of English-language posts. Topic modeling is one of the most useful machine learning tools for analyzing large collections of unstructured data, especially large corpora of text documents. LDA is used to decompose documents into latent components, or topics, where each topic is a distribution over words intended to capture the thematic structure of both the corpus and the individual documents themselves. The observed variables %on the other hand 
%are words in each document: each document is assumed to be a bag-of-words, where the order in which words appear does not matter. 

% In LDA, each document is assumed to be a mixture of the topics, and is drawn from the following generative process:

%  \begin{enumerate}
%      \item Choose topic proportions $\Theta \sim  Dirichlet(\alpha)$ 
%      \item For each word $w_{d,n}$ in document d:
%      \begin{enumerate}
%      \item Choose a topic $z_{n} \sim Multinomial(\Theta)$
%      \item Choose a word $w_{d,n} \sim Multinomial(z_{n})$
%      \end{enumerate}
%  \end{enumerate}

%Once we train the model we can represent each document as a distribution over its constituent topics. Doing so gives us a low-dimensional representation of the document. %which we can then average over all the documents in a forum to get a topic distribution for that forum. 

%\note{todo: Nathan}
%\note{Should software citations be footnotes instead?}

In this paper we use the Gensim implementation of LDA \cite{rehurek_lrec} to learn a model with 100 topics. We train the model on all 1.33 million documents to learn the most informative topics. We tested with 50, 100, and 200 topics respectively, and found that 100 topics results in the most coherent and relevant topics. To examine the dynamics, we focused on time period of 2016 until mid September 2017 which has the best coverage in our data. We only looked at forums with at least one month of activity and more than 100 posts overall, which reduced our data set to 80 forums (and approximately 482 thousand posts). Figure \ref{fig:activity} shows the level of activity in these 80 forums. The activity is highly heterogeneous, with some forums seeing hundreds of posts weekly, and other forums showing little activity. %qualitatively

%For this study, we focus on documents categorized as English-language posts. 

%We hold the following parameters constant during training: we make one pass over the corpus, evaluating once every 1.2 million documents, using a symmetric alpha and eta and iterate 1500 times (or until convergence at 0.001). %Based on perplexity computed over a heldout set of 2449 documents as reported in XXX, we select the model with 200 topics as having the lowest perplexity. 
% \begin{table}[]
%     \centering\small
%     \begin{tabular}{|c|c|}
% \hline
%      \# Topics & Perplexity \\
%      \hline
%     50  & 494.1 \\
%     100 & 503.8 \\
%     200 & \textbf{447.8}\\
%     \hline
% \end{tabular}
%     \caption{Confusion matrix for 50, 100 topics
% }
%     \label{tab:confusion_matrix}
% \end{table}
% %Nathan -- what are we evaluating here with the confusion matrix

\begin{figure}
%\vspace{14pc}
\includegraphics[width=0.9\columnwidth]{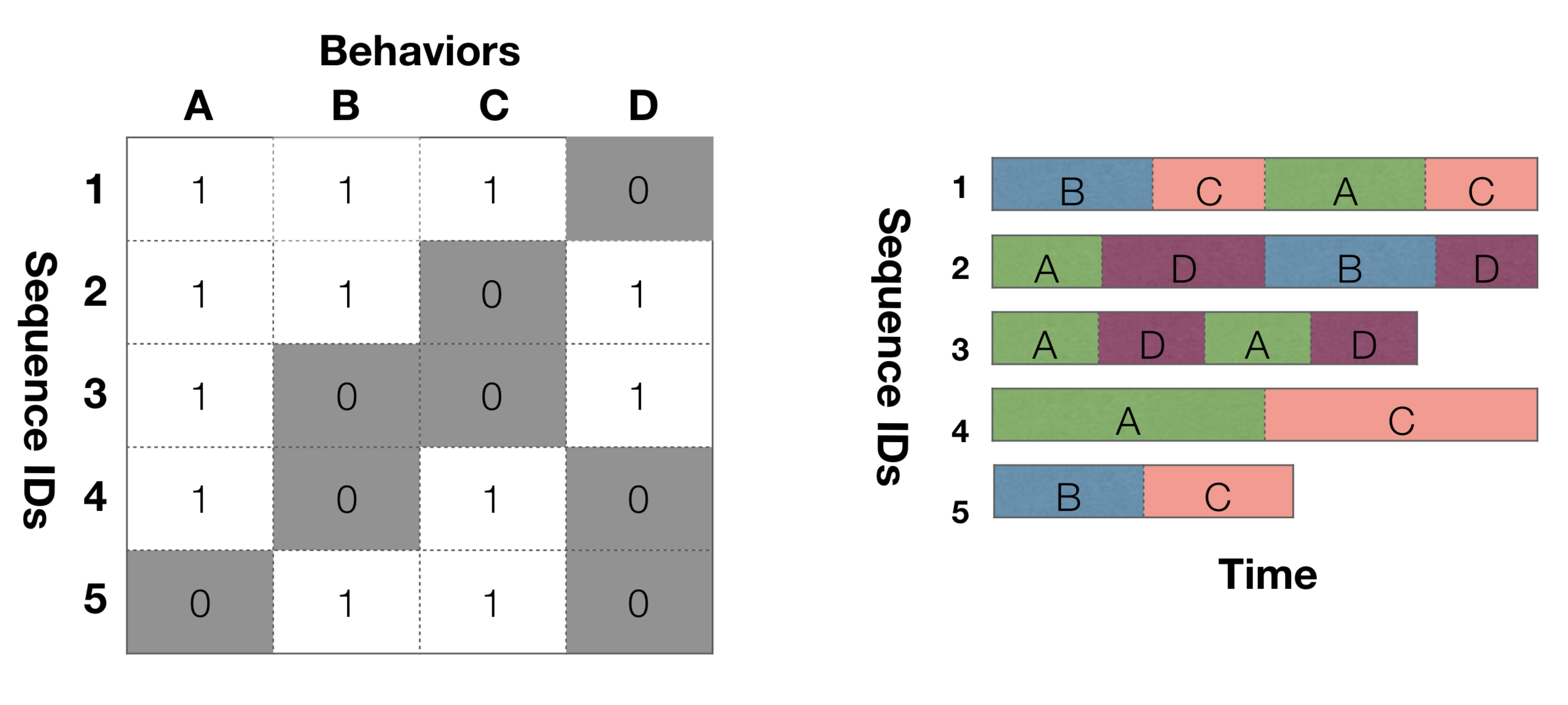}
\caption{Illustration of the model. Left image is matrix $F$. Based on this matrix time series $1$ exhibits states A,B and C but doesn't exhibit state D. The right image shows segmentation of time series with hidden states}
\label{fig:bphmm}
\end{figure}

\subsection{Modeling Dynamics of Activity}
\label{sec:dynamics}
%Beta Process HMM}
%One of the most popular tools for studying multivariate time series are the vector autoregressive (VAR) models~\cite{monbet2017sparse}. In a VAR model of lag $r$, each variable is a linear function of itself and the other variable's $r$ previous values.  %$r$ previous values. \note{Such models take as input a set of time series and produces ...} 
%However, such models cannot describe time series with changing behaviors. %, for example cases where behavior switches between different states.
%In order to model such cases, we use Markov switching autoregressive models, which are generalization of autoregressive and Hidden Markov Models~\cite{monbet2017sparse}. 

Hidden Markov Models (HMMs) have been used extensively in modeling dynamic processes and time series in a variety of applications. These generative models segment time series into a predefined number of latent states and learn transition rates between them.

When modeling multiple dynamic processes with HMMs, it is useful to represent them with global states that are shared between these time series, rather than modeling each dynamic process independently and then learning the mapping between states. %\note{cite example}. 
Joint modeling facilitates comparison of %the dynamics of 
different processes and learns more generalizable models. It is also convenient to work with a non-parametric model that does not fix the number of states a priori. %and nature 

In this paper, to model activity within the d2web, we describe each forum as a time series of topic vectors representing discussions. We use a generative model  proposed by \cite{fox2014joint,fox2009sharing}, called \textit{Beta Process HMM} (BP-HMM), to identify latent states shared by different time series. % representing the  forums. 
  Based on the proposed model, different time series are described by a subset of shared latent states. The states are represented by a binary matrix $F$, where $F_{ij} = 1$ means time series $i$ is associated with state $j$. An example of $F$ matrix is shown in the left panel of Figure ~\ref{fig:bphmm}. Given matrix $F$, each time series is modeled as a separate hidden Markov model with the states it exhibits. Each global state is modeled using a multivariate Gaussian distribution, when a time series is in state $x$, its data is sampled from a Gaussian distribution with mean vector $\mu_x$ and covariance matrix $\Sigma_x$. %\note{AA: Nazgol I changed it to multivariate Gaussian, please confirm is correct. }% An HMM is represented by a transition matrix $T_i$, which is a square matrix with dimensions equal to the number of states. %In the context of this paper, states represent behaviors that the time series $i$ exhibits.
%$K_i = \sum_k F_{ik}$
%The entry $T_i(m,n)$ gives the probability that a time series $i$  transitions from state $m$ to state $n$. The matrix $T_i$ is stochastic, with the sum of entries in each row equal to $1$. 
%Rows of transition matrices, transition distributions $\pi_{k,i}$, %for each state 
%are sampled from a finite Dirichlet prior. %with dimension equal to number of behaviors a particular time series exhibits.

%Having the transition matrices, time series $i$ is represented by a sequence of states, %$\text{seq}_i = 
%$[z_1 \cdots z_T]$; therefore, $z_t$ represents the state of the time series $i$ at time $t$.
 %vector autoregressive process with lag $r$.
%\begin{align}
%\begin{split}
%    y_t = \sum_{l=1}^{r} A_{l,z_t}y_{t - 1} + e_{z_t}
%    \\
%    e_{z_t} \sim\ \mathcal{N}(0,\sigma^{2}_{z_t})
%\end{split}
%\end{align}
%Conditioned on latent state at time $t$ $z_t$,
%When a time series is in state $z_t$, it's future values evolve according to the autoregressive weights $A_{1,z_t} \cdots A_{r,z_t}$ and noise $e(z_t)$.

Since the number of such states in the data is not known a priori, the Beta process is used~\cite{hjort1990nonparametric,thibaux2007hierarchical} as a prior on matrix $F$. A Beta process allows for infinite number of states but encourages sparse representations. Consider, as an example, a model with $K$ states. Each state (column of matrix $F$) is modeled by a Bernoulli random variable whose parameter is obtained from a Beta distribution (Beta Bernoulli process), i.e.,
\begin{equation}
\begin{split}
\theta_k \sim\ \text{Beta}(\alpha/k,1), k = 1,\cdots,K 
\\
 F_{nk} \sim\ \text{Bernoulli}(\theta_k), n = 1,\cdots,N
\end{split}
\end{equation}

The underlying distribution when this process is extended to an infinite number of states, as $K$ tends to infinity, is the Beta process. This process is also known as the \textit{Indian Buffet Process} \cite{ghahramani2006infinite,thibaux2007hierarchical}
%\textit{Indian Buffet Process} and the underlying distribution is the Beta process%, which is an instance of completely random measure 
%~\cite{kingman1967completely,griffiths2011indian}. The Indian Buffet Process allows to generate distributions over binary matrices with an infinite number of columns (states) and finite number of rows (time series).
which can be best understood with the following metaphor involving a sequence of customers (time series) selecting dishes (states) from an infinitely large buffet. 
The first customer enters the buffet and selects servings from $Poisson(\alpha)$ number of dishes. The $n$-th customer selects dish $k$ with probability $m_k/n$, where $m_k$ is the popularity of the dish, yielding the so-called ``rich-get-richer'' effect, and $Poisson(\alpha/n)$ new dishes.%i.e., some features are going to be more prevalent than others. S/He then selects $Poisson(\alpha/n)$ new dishes.      
 
With this approach, the number of states can grow arbitrarily with the size $n$ of the dataset: in other words, the number of states increases if the data cannot be faithfully represented with the already defined states. However, the probability of adding new states decreases according to a $Poisson(\alpha/n)$.  Finally, the distribution generated by the Indian Buffet Process is independent of the order of the customers (time series). For posterior computations based on MCMC algorithms, the original work is referenced \cite{fox2014joint,fox2009sharing}.

%where P and Q refer to probability distributions, in this case, two multinomial distributions over topics. 

%In this paper, we used the states learned through this framework for %clustering, classification, and regression of human-generated biometric signals captured from wearable sensors. 
\begin{figure}
    \centering
    \begin{tabular}{@{}c@{}c@{}}
    \includegraphics[width=0.45\columnwidth]{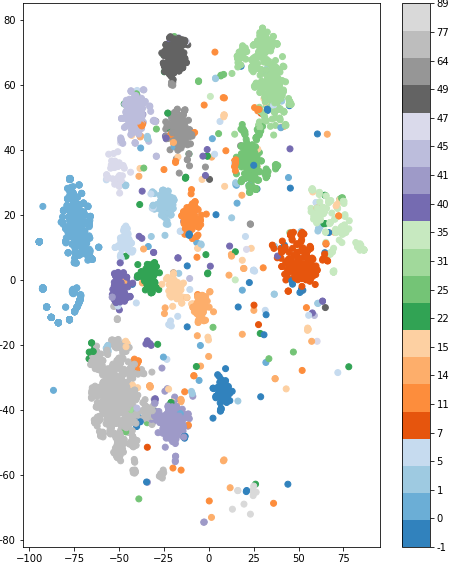}
    &
    \includegraphics[width=0.45\columnwidth]{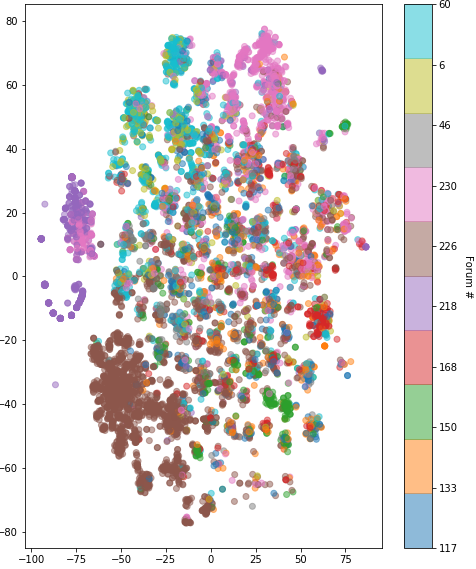}
    \\ (a) topics & (b) forums\\
    \end{tabular}
    \caption{Visualization of the d2web discussions in May 2016 using t-SNE. Each dot represents a post, with its color representing (a) the topic or (b) forum to which the post belongs. The top 20 topics  learned by the 100-topic LDA are shown, and the post was assigned to the highest probability topic. }
    \label{fig:tsne_topics}
\end{figure}

\subsection{Clustering}

\begin{table*}
\centering
\begin{tabular}{ l | l }
\hline
Topic (Manually Labeled) & Top 10 Keywords \\
\hline
\textbf{Vending}\\
\hspace{3mm}1. Locations & checked, live, united states, unknown, california, carolina, south, ca, nj, new \\
\hspace{3mm}2. Money & money, people, make, pay, want, free, buy, just, like, sell \\
\hspace{3mm}3. Pharmaceuticals & buy, online, prescription, cheap, cod, xanax, delivery, overnight, order, day \\
\hspace{3mm}4. Banking & card, bank, credit, cards, paypal, account, business, debit, accounts, gift \\
\hspace{3mm}5. Fake IDs & fake, real, id, original, high, english, license, quality, registered, passports \\
\hspace{3mm}6. Purchase details & order, vendor, days, sent, orders, ordered, package, received, shipped, just \\
\hspace{3mm}7. LSD & like, just, quote, lsd, really, good, feel, know, tabs, experience \\
\hspace{3mm}8. Cryptocurrency & bitcoin, btc, wallet, address, send, coins, bitcoins, transaction, sent, account \\
\hspace{3mm}9. Marijuana & like, good, got, weed, time, bit, high, great, nice, low \\
\hspace{3mm}10. Markets & market, vendor, vendors, dream, alphabay, markets, scam, ab, hansa, escrow \\
\hspace{3mm}11. Narcotics & good, cocaine, quality, best, vendor, product, mdma, coke, free, order \\
\textbf{Security}\\
\hspace{3mm}12. Malware & virus, scan, antivirus, file, malware, clean, av, security, download, detected \\
\hspace{3mm}13. Botnets & bot, attack, malware, used, domain, ddos, botnet, irc, hosting, attacks \\
\hspace{3mm}14. Windows & windows, build, microsoft, xp, vista, beta, server, ms, longhorn, version \\
\hspace{3mm}15. Social hacking & email, send, rat, stealer, message, keylogger, mail, facebook, crypter, download \\
\hspace{3mm}16. Law enforcement & police, law, drug, drugs, enforcement, according, dark, illegal, said, darknet \\
\hspace{3mm}17. Hacking tutorial & learn, know, want, good, learning, start, like, knowledge, programming, hacking\\ 
\hspace{3mm}18. Carding & transfer, dumps, info, sell, cvv, good, track, balance, bank, uk \\
\hspace{3mm}19. Web Vulnerabilities & web, sql, php, injection, exploit, code, server, site, script, page \\
\hspace{3mm}20. OS Code & process, code, dll, memory, address, api, function, module, use, hook\\
\hspace{3mm}21. Network Hacking & network, connect, wifi, wireless, ip, internet, router, connected, pineapple, fon \\
\hspace{3mm}22. Security & information, data, security, software, used, access, user, users, network, application \\
\hspace{3mm}23. Proxy & use, tor, using, vpn, internet, proxy, browser, ip, web, access \\
\hspace{3mm}24. Mobile phones & phone, android, phones, samsung, pixel, battery, note, camera, better, google \\
\hspace{3mm}25. Update & install, installed, download, just, update, need, use, installing, using, try \\
\textbf{Gaming}\\
\hspace{3mm}26. Gaming Source Code & end, local, return, function, false, mod, script, item, nil, damage \\
\hspace{3mm}27. Torrents & torrent, quote, download, upload, forget, left, feedback, like, plz, 720p \\
\hspace{3mm}28. Gameplay & game, complete, level, win, play, kill, mode, team, player, single \\
\hspace{3mm}29. Games & game, games, new, play, like, xbox, ps4, sony, playstation, console \\
\hspace{3mm}30. PlayStation Vita & vita, ps, psp, game, firmware, exploit, games, sony, custom, psn \\
\hspace{3mm}31. Emulators & game, games, vita, version, plugin, homebrew, psvita, emulator, use, play \\
\hspace{3mm}32. Hacking Consoles& ps3, games, play, tutorial, game, cfw, console, psn, use, need \\
\textbf{Other}\\
\hspace{3mm}33. Contact & contact, pm, need, icq, want, send, add, rue, interested, na \\
\hspace{3mm}34. Thanks & thanks, thank, man, lot, sharing, bro, thx, share, mate, nice \\
\hline
\end{tabular}
\caption{Example topics in the 100-topic LDA model.Topics are labeled (in the first column) manually for convenience. }
\label{tab:interesting_topics_english}
\end{table*}

\label{sec:clustering}

%When applied to multivariate biometric signals, the  %inference process
%generative model described above learns a hidden Markov model representation of each multivariate signal. We can use the learned HMMs to identify individuals with similar behaviors. In order to do this, we need the ability to cluster similar HMMs.
%hence their transition matrix could be viewed as generative models
To define a similarity measure between two HMMs, one could measure the probability of their state sequences %, $\text{seq}_i = 
%$[z_1 \cdots z_T]$, %\note{$\text{seq}_j$(?)}  
having been generated by the same process. Since  each time series is associated with  a distinct generative process, we measure two state sequences' %their \note{Who is ''they'': sequences? Or models?} %similarity as the likelihood that $\text{seq}_i$ was generated by a process or transition matrix of time series $j$ and likelihood of $\text{seq}_j$ coming from transition matrix $i$ and since we wanted a symmetric distance we averaged the two computed values.
similarity as the likelihood that $\text{seq}_i$ was generated by the process that gave rise to $\text{seq}_j$,  and the likelihood that $\text{seq}_j$ was generated by the process giving rise to $\text{seq}_i$. We average the two likelihoods to symmetrize the similarity measure.

\begin{equation}
\forall_{i,j} \text{Sim}(i,j) = \frac{p(\text{seq}_i|T_j) + p(\text{seq}_j|T_i)}{2}
\end{equation}

The likelihood $p(\text{seq}_i|T_j)$ is computed using the learned transition matrix $T_i$ and Markov process assumption. In transition matrix $T_i$, which is a square matrix with dimensions equal to the number of states, entry $T_i(m,n)$ gives the probability that time series $i$  transitions from state $m$ to state $n$. Matrix $T_i$ is stochastic, with the sum of entries in each row equal to $1$.
%\begin{align}
%\begin{split}
%&T_i = \begin{bmatrix} \pi_{1,i} \\ \vdots \\ \pi_{K,i} %\end{bmatrix} \\
%& z_t|z_{t-1} \sim\ \pi_{z_{t-1},i}\\
%\end{split}
%\end{align}
Once the similarity between two HMMs is defined, we can perform a number of operations, including clustering similar time series together. For example, we use hierarchical agglomerative clustering method to automatically group forums (represented by their time series) with similar discussions.
%was used to understand the structure of participants in the study. ....

%\Andres{AA: I don't see this as a problem, specially given tat you have a symmetric similarity measure: You only care about  what is the best model for a sequence. True, for larger sequences the likelihood is smaller, but this is across all possible models. }
%As a side note for time series with different lengths, longer time series will automatically have a smaller likelihood, since they are multiplied by a number smaller than $1$ more times. To solve this issue we normalize the likelihoods based on the length of the time series %or 
%lengthen shorter time series to the max length using their generative process, 
%then compute their similarities. %For this paper 
%We normalized the likelihoods by dividing them to ${\frac{1}{K}}^L$, $K$ being the number of states and $L$ being the length of the time series. % of $p(\text{seq}_i|T_j)$ 
%Here $\pi_{k,i}$ is state specific transition distribution of time series $i$ and $T_i$ is it's transition matrix.
%Sim($i,j$) is the average of the likelihood of sequence $i$ coming from transition matrix $j$ and sequence $j$ coming from transition matrix $i$.

\begin{figure*}%[h!]
%\vspace{14pc}
\centering
\includegraphics[width=\linewidth]{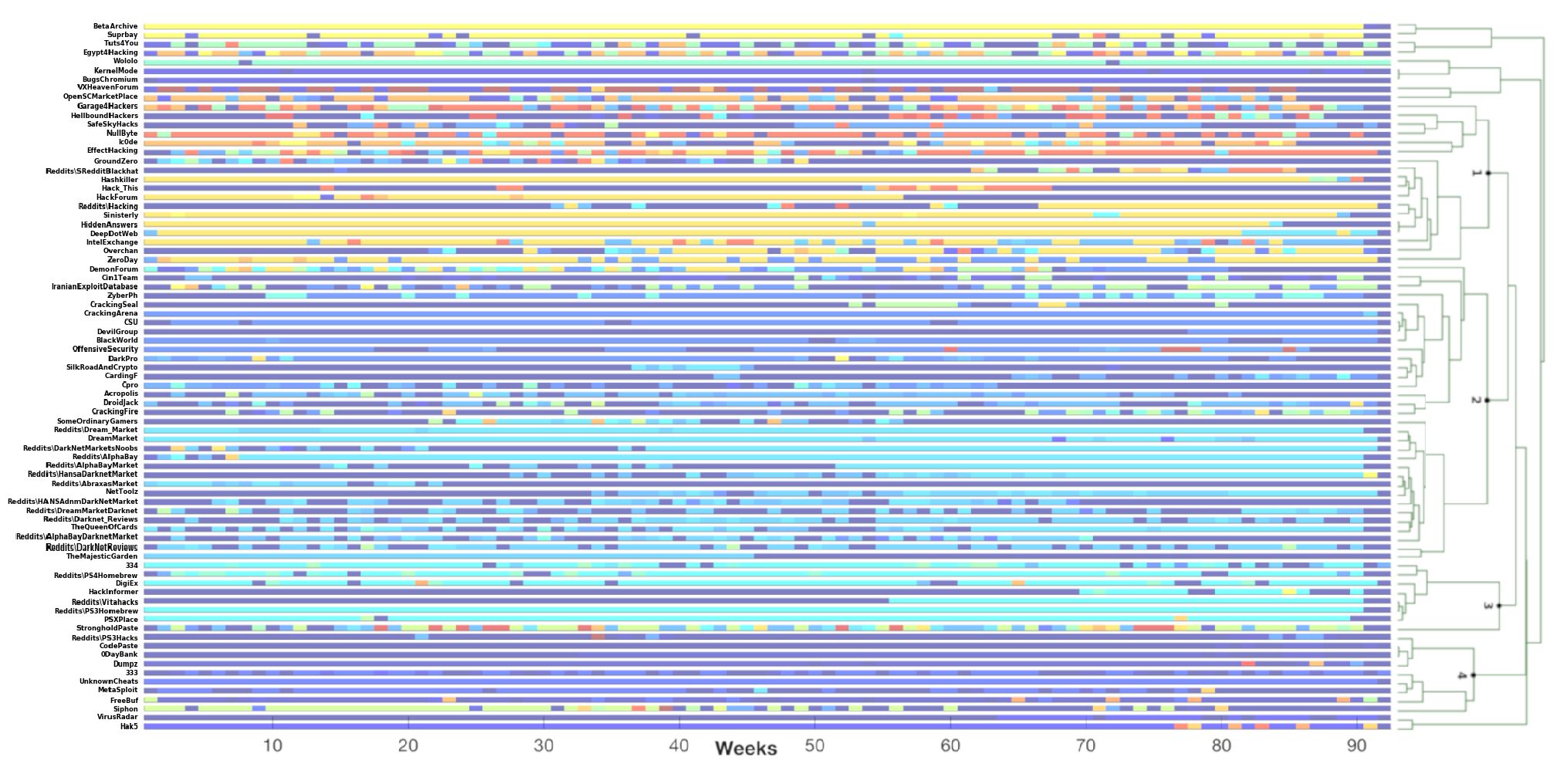}
\caption{State sequences of forums (each color represents a state) and Dendrogram showing the similarity of forums based on their learned states. }
\label{fig:clustering}
\end{figure*}

\section{Results}

\subsection{Topic Analysis}

In this section we explore topics learned by Latent Dirichlet Allocation (LDA) by looking at their most significant words. %from each of the 100 topics in the model and report 5 of the most interesting topics. 
There are a wide variety of topics covered in our dataset. %in the 100 topics %including the ones in 
 Table~\ref{tab:interesting_topics_english} highlights some of the topics learned by the 100-topic LDA model by showing significant words associated with them. %, other topics include % and more including 
We visualized the topics using t-distributed Stochastic Neighbor Embedding (t-SNE) \cite{maaten2008visualizing} in Figure~\ref{fig:tsne_topics}(a). Each dot on the plot represents a post, colored by its most important topic. Alternatively, we can color the posts in the t-SNE space by the forum on which they were posted (Figure~\ref{fig:tsne_topics}(b)). The clustering of posts in the forum space suggests that topics are highly concentrated around specific forums.%, and that conditioned on the time interval we look at, topics would be able to distinguish forums.  Posts clustering together in the topic space indicates significant semantic separation between topics

\subsection{Forum Dynamics}

%Forum Dynamics
%Given a time interval (day may be too sparse, so week or month), represent each forum as a time series of the topic vector of posts aggregated over that time interval. Visualize the dynamics over time of 2-3 most active forums. Are they on-topic, or do topics fluctuate frequently? 

%Method1: Compute cross-entropy between two consecutive time intervals for a forum (topic vector at time $t$ with topic vector at $t+1$)

%Method 2: Compute average topic vector $T_{ave}$, then at each time $t$, compute cross entropy for of topic vector at $t$ with $T_{ave}$. Plot these values ove

%For training BP-HMM  

%\begin{figure*}[tbh]
%\vspace{14pc}
%\centering
%\includegraphics[width=0.95\linewidth]{images/DW_State_seq}
%\caption{State sequences of forums (each color represents a state) ordered from top to bottom with the same ordering of plot ~\ref{fig:clustering} from left to right }
%\label{fig:StateSeq}
%\end{figure*}

To examine the dynamics of topics on d2web forums, we represent each forum as a time series of topic vectors learned by the 100-topic LDA model. Our unit of time in this analysis is a week; to generate a forum's vector we average the topic vectors of all posts submitted to the forum over the course of a week. The time series of weekly topic vectors were used to both learn HMM states and compute cross-entropy (cross-entropy is discussed further in \ref{cross_intropy}).

%For further analysis, we focused on time period of 2016 until mid September 2017 which has the best coverage. Also we only looked at forums with at least one month  of activity and more than 100 posts overall, which reduced our data set to 80 forums (and approximately 482 thousand posts). Figure \ref{fig:activity} shows the level of activity in these 80 forums. The activity is highly heterogeneous, with some forums seeing hundreds of posts weekly, and other forums showing little activity. 

After training the BP-HMM model on weekly topic distributions of forums the model learned 28 states, i.e., 28 different topic distributions. We clustered the forums according to the similarity of their learned states using the method described in Section~\ref{sec:clustering}. Figure~\ref{fig:clustering} shows the resulting dendrogram and also the sequences of learned states for each forum. Each line in the figure represents a forum, and different states are represented by different colors. Transitions between states are visible in places the colors alternate.
%For clarify of visualization, the forums are in the same order as shown in the hierarchical clustering dendrogram. Nearby forums exhibit similar color schemes and therefore similar activity patterns.

%\subsubsection{Analyzing Clusters}
The clustering results show that the method is able to cluster forums into meaningful groups. Next we examine a few of the main clusters:

\textbf{Cluster 1} mostly contains forums discussing cyber hacking, including HackForum, GroundZero, ZeroDay, DeepDotWeb, SafeSkyHacks. The two subgroups in this cluster differ mainly in their levels of activity. Forums in the first subgroup are less active (5 posts in a week on average), which is why their average weekly topic vector is more sensitive and their corresponding HMM changes state more frequently. In the active group (more than 50 posts in a week on average) the most common state is the yellow state, which corresponds to activation of the following topics (described in Table~\ref{tab:interesting_topics_english}): 16.Law enforcement,  23.Proxy and 17.Hacking tutorial.

\textbf{Cluster 2} mostly contains dark web marketplaces such as Abraxas Market, AlphaBay, Dream Market, Hansa Market and BlackWorld, as well as forums dedicated to their reviews. This cluster is also divided into two main subgroups: in the first subgroup the dark blue state dominates representing high activity of discussions regarding topics 1.Locations (which is mostly concerned with the sale of proxy servers), 33.Contact, 34.Thanks and 4.Banking. The second subgroup, depicted with light blue states, corresponds to the topics 6.Purchase details, 10.Markets, 8.Cryptocurrency and 11.Narcotics. Based on the clustering, one can characterize forums in the first subgroup as mostly selling proxy servers and sharing information about other marketplaces,  while forums in the second subgroup are more involved in selling drugs. In section~\ref{ProxyServers} we take a deeper look into forums in the first subgroup.

\textbf{Cluster 3} is made up of forums related to hacking Playstation video game consoles where the most prominent state in this cluster is the Cyan state. The most active topics in this state are 32.Hacking Consoles and 25.Update.% from table~\ref{tab:interesting_topics_english}.

\textbf{Cluster 4} (as well as the two forums adjacent) contains forums which are predominantly focused on white hat hacking. Notable forums include Metasploit and Hak5. While 0daybank and FreeBuf are related, they are mostly in Chinese, hence their more active topics have many non-English tokens and are hard to interpret.
% The yellow cluster represents forums related to dark web marketplaces, such as Abraxas Market, Alpha Bay, Dream Market, Hansa Market, as well as forums dedicated to their reviews. The nearby cluster marked in black contains forums related to hacking Playstation.

% \paragraph{volatility of forums}
%Next, we measure the volatility of forums using the HMM representation as a complementary measure to the results obtained with cross entropy over the LDA topics. 
With the state sequences obtained from our BP-HMM model, we can track forums' discussions. State transitions indicate a significant change in discussions and could represent an event. However, as shown in Figure~\ref{fig:clustering}, some forums change states more frequently, thus their transitions might have less significance. In order to be able to recognize significant transitions, we describe the volatility measure, a forum's likelihood to drastically change its own topic distribution. As each forum is characterized by its learned transition matrix over the global states, we compute a forum's volatility by %can get an estimate of that forum's likelihood to change state, which in turns means a drastic change of the topic distribution. Specifically, 
%for each forum by 
adding the off diagonal elements of its transition matrix, the probability of changing states. Since we are interested in finding variations in topics discussed rather than the activity of forums, the probabilities of the state corresponding to $0$ posts (i.e., no data) were not taken into account. Also to validate the results obtained, a similar volatility measure was computed with cross entropy and is described in section \ref{cross_intropy}. Table ~\ref{tab:volatility} shows a list of forums with high and low volatility computed via both methods.
%\note{AA: is this correlation (volatility with activity) true only considering the highly active forums? or just due to the inactive ones?}
%As seen in the state sequences and clustering, a forum's volatility is correlated with its activity level as some of the most volatile forums are forums with low activity. Again, this makes sense since their weekly topic vectors get averaged over fewer posts and are more susceptible to noise.  

Using the described volatility measure with the HMM, we find that the most volatile forums with at least 10 posts per week (on average) are OpenSC Marketplace, Stronghold Paste and Demon Forum and that the least volatile forums are BugsChromium, CSU, and the subreddit PS3Homebrew. An estimate of forum's volatility or lack thereof is also apparent from its state sequence in Figure~\ref{fig:clustering}. Stronghold Paste is an onion website similar to Pastebin and covers different topics and hence different states. However, there are two main states it oscillates between: in one of them hacking and cyber security topics (described by topic 19.Web Vulnerabilities%from Table \ref{tab:interesting_topics_english}
) are more prominent, and in the other one, topic 8.Cryptocurrencies has %higher rates.
high activation. These results show state transitions in forums like Stronghold Paste have a low probability of being indicative of an event. However transitions in forums like BugsChromium or CSU might be of more interest. %On the other hand forums such as CSU are more focused on specific topics.  %\note{Comparing results from cross entropy and HMM} 

\subsection{Topic Dynamics}\label{cross_intropy}

 %Also some forums have high level of activity for a small period of time and very low activity for the rest of the study, for example 

Since HMM segments forum dynamics into discrete states, it is less sensitive to noise observed in the data. On the other hand it might not be able to capture small meaningful changes or trends. Hence, we also compute cross entropy of forums as a measure of dispersion over time to validate results obtained with non-parametric HMM model. To compute the cross-entropy we use the following formula where Q is the topic distribution for a forum averaged over the entire timespan and P is the topic distribution in a unit of time.
\begin{equation}
     H(p,q) = E_{P}[- \log_2 Q(x) ]
    %$H(p,q) = - \sum_{x \in X} p(x)\log_2 q(x)$
\end{equation}
%We computed cross-entropy per Section \ref{sec:dynamics} and examined the entire corpus. 
%In Fig. \ref{fig:cross-entropy-all-forums} we can see that 
The majority of the forums have cross-entropy values in the same range. %, between 3 and 4.5.
The forum with the lowest average cross-entropy (and therefore lowest volatility)
%and the five forums with the most activity in  
is CSU, a forum primarily concerned with credit card dumps, and as such makes sense that the forum would focus on very few topics of discussion. %(Tab.~\ref{tab:volatility}).
On the other hand, forums like CodePaste and DroidJack have some of the highest values of cross-entropy, which suggests that the topics of discussion are dispersed and change more often in these forums. Table~\ref{tab:volatility} shows forums with lowest and highest average values of cross-entropy, used as a measure of volatility.
%\note{AA: The following seems inconsistent with the volatile results from before }
This is consistent with the result we got from the HMM model, in the sense that forums with low or high volatility based on the HMM measure also appear in the bottom or top of the ranking based on the cross entropy measure. Results show that large and active forums have wide-ranging discussions on diverse topics however their average topic distribution is usually consistent over time. Forums focused on specific topics like CSU also tend to have low volatility. On the other end of the spectrum there are forums with medium or low activity with discussions spanning a wide range of topics like Stronghold Paste and CodePaste. %It makes sense that we see slightly different orders between the cross-entropy and HMM rankings since the cross-entropy computation is smoothed by a rolling average, potentially nullifying a major change in value that would shift the mean.  %spanning hacking, games, and marketplaces. Likewise, with code-sharing sites, we see a number of different kinds of source-code and natural language posts that span multiple different topics (and sometimes mix languages as well). 
\begin{table}%[]
\centering
\begin{tabular}{  c | c }
\hline
 HMM-ranked volatility & Cross-entropy-ranked volatility\\
\hline

\hspace{3mm}1. GroundZero & CodePaste \\
\hspace{3mm}2. OpenSCMarketPlace & DroidJack \\
\hspace{3mm}3. StrongholdPaste & EffectHacking \\
\hspace{3mm}4. DemonForum & DemonForum \\
\hspace{3mm}5. EffectHacking & Overchan \\
\hspace{3mm}6. CrackingFire & HellboundHackers \\
\hspace{3mm}7. Overchan & CardingF \\
\hspace{3mm}8. NullByte & HackForum \\
\hspace{3mm}9. Siphon & Reddits$\setminus$ Hacking \\
%\hspace{3mm}10. Garage4Hackers & Sinisterly\\
 ... & ...\\
\hspace{3mm}71. UnknownCheats & TheMajesticGarden \\
\hspace{3mm}72. DevilGroup & Dumpz \\
\hspace{3mm}73. KernelMode & KernelMode \\
\hspace{3mm}74. VirusRadar & BlackWorld \\
\hspace{3mm}75. Reddits$\setminus$Vitahacks & MetaSploit \\
\hspace{3mm}76. BetaArchive & 0DayBank \\
\hspace{3mm}77. TheMajesticGarden & Reddits$\setminus$DarkNetReviews \\
\hspace{3mm}78. Reddits$\setminus$PS3Homebrew & Wololo \\
\hspace{3mm}79. CSU & VirusRadar \\
\hspace{3mm}80. BugsChromium & CSU \\
\hline
\end{tabular}
\caption{Volatility computed via HMM and via Cross-entropy. Ranked highest volatility to lowest.}
\label{tab:volatility}
\end{table}
 %but mostly hacking and cyber security topics like topic 19.Web Vulnerabilities 

\subsection{Case Studies}
\subsubsection{Prescription Drugs}
In this section we give an example of how this framework could be used to study d2web discussions. In the search for anomalies using the results obtained by HMM we observed a rare state, exhibited only on 3 first weeks of June 2017 and first week of August 2017  by forum OffensiveSecurity where topic 3.Pharmaceuticals % as reported in Table \ref{tab:interesting_topics_english}
has its highest probability. OffensiveSecurity is considered a low volatility forum based on both measures computed in Table \ref{tab:volatility} which makes this transition more of interest. %By looking at the posts published on this forum and on the aforementioned dates we retrieved approximately 100 documents with the average probability of above 0.90 for topic 3.Pharmaceuticals.    % The second case study is the discussion of the sale and/or acquisition of prescription drugs. Here we demonstrate that we can use a specific state from the HMM model to identify high activity related to pharmaceuticals. 
%\begin{figure}
%    \centering
%    \includegraphics[width=\columnwidth]{images/topic_6_prescriptions.png}
%    \caption{Distribution of prescription-drug topic over the timespan of the corpus. Each %point is the average probability of that topic for a forum for that week.}
%    \label{fig:prescription-topic}
%\end{figure}

%\begin{figure}
%    \centering
%    \includegraphics[width=\columnwidth]{images/topic_91_drugs.png}
%    \caption{Distribution of illicit-drug topic over the timespan of the corpus. Each point is the average probability of that topic for a forum for that week.}
%    \label{fig:drug-topic}
%\end{figure}

%\begin{figure}
%    \centering
%    \includegraphics[width=\columnwidth]{images/topic_91_drugs.png}
%    \caption{Boxplot Distribution of 3.Pharmaceuticals topic over the timespan of the corpus. Each point is the average probability of that topic for a forum for that week.}
%    \label{fig:prescription-topic}
%\end{figure}
% Over the course of 4 days the same kind of post was posted,
By looking at the posts published on this forum and on the aforementioned dates we retrieved similar posts
with variations in the names of the drugs being advertised. An excerpt from one of the posts is as follows: \say{...buy  lynoral cheap buy generic  femara buy  modafinil online uk can i buy  qsymia online buy  cytotec online us buy  lumigan online canada buy 25 mg  lyrica buy vibramycin florida buy  diflucan without buy  synthroid online next day delivery buy  xanax online us where to buy generic  qsymia buy  adderall no prescription buy  cytotec in europe...}

This analysis shows a high and anomalous volume of advertisements for prescription drugs in the specified dates which suggests some precipitating event that merits further investigation.
%. When we examine the forum-weeks with an average probability in the topic over 0.50, we return approximately 100 documents from June 2017 from OffensiveSecurity (average > 0.90). 
%Other forum-weeks that have less significant average probabilities than OffensiveSecurity, describe more discreet conversations about prescription drugs on other forums, most notably the Dream\_Market subreddit and DreamMarket forum itself.  
%In Fig. ~\ref{fig:prescription-topic} we can see that there are few forums that have probability over 0.50 at any point in time in the corpus. Further inspection of the sole point in June 2017 reveals around 100 posts that seem to be spam from Forum 84. Other points do seem to suggest more discreet conversations about prescription drugs on other forums, most notably Forum 17 and 60, forums which are mostly focused on DreamMarket.
%\note{nathan: do we need to keep saying as reported in Table ???}

We also looked at one of the most illicit drug topics, 11.Narcotics. We found that besides 11.Narcotics other prominent topics in the state where this topic is at its highest value are 9. Marijuana and 7.LSD. These are a few forums which exhibit this state: The Majestic Garden, the DarkNetReviews subreddit, and the HANSAdnmDarkNetMarket subreddit.

%When we look at one of the most prominent illicit drug topics, 11.Narcotics as reported in Table \ref{tab:interesting_topics_english}, we recover approximately 1000 documents from DreamMarket itself, and 60 documents from two subreddits, AlphaBay and AlphaBayMarket. Of note are 75 documents from the NetToolz forum, where vendors seem to be introducing themselves in a templated manner. 
\subsubsection{Proxy Servers}\label{ProxyServers}

\begin{figure}
   \centering
    \includegraphics[width=\columnwidth]{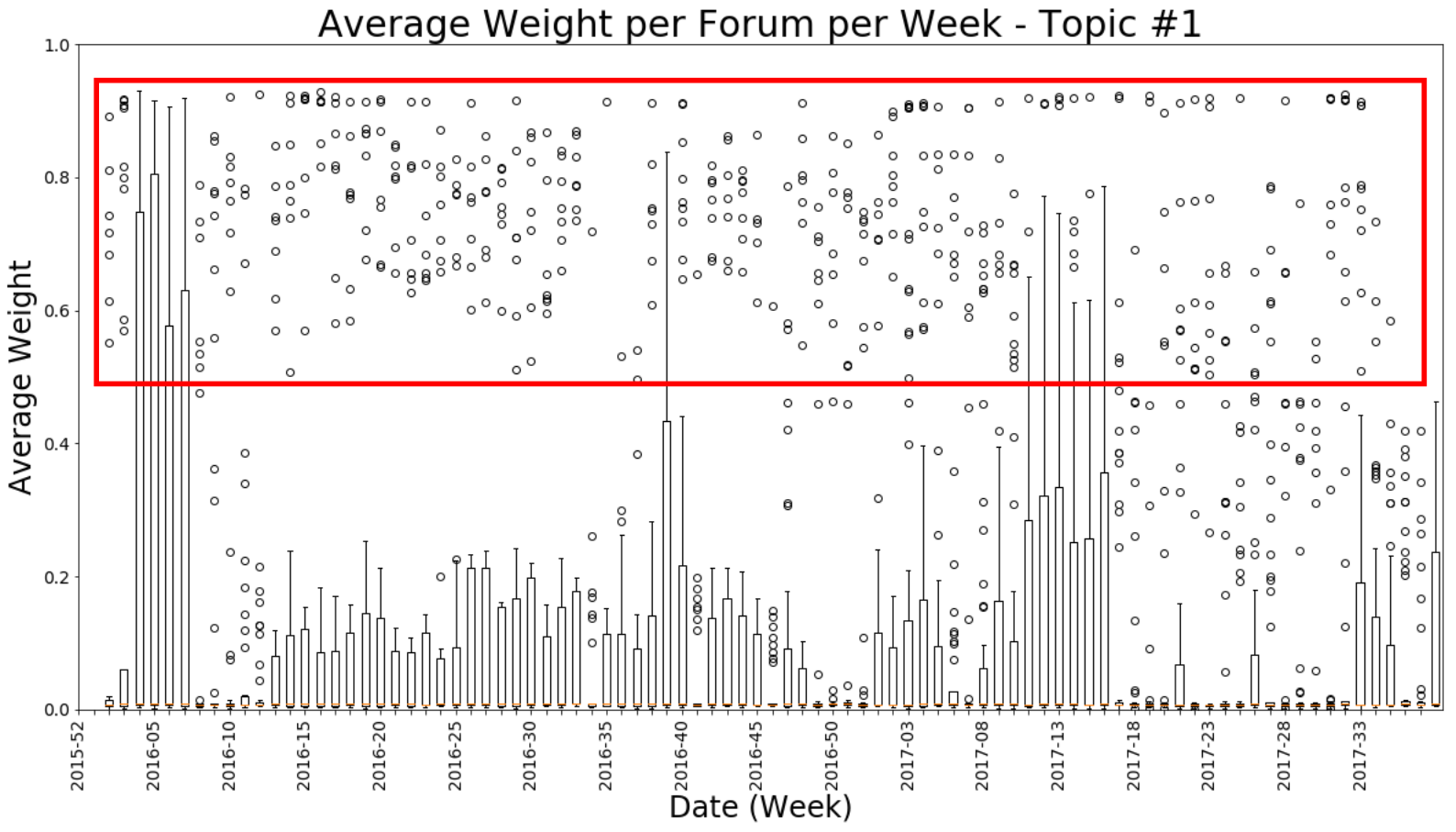}
    \caption{ Boxplot Distribution of Proxy-server topic over the timespan of the corpus. Plotted are the 10 top forums in Cluster 2. Each point is the average weight of that topic for a forum for that week. The red box indicates the significant forums in each week.
    }
    \label{fig:proxy-topic}
\end{figure}

The second case study is around the discussion of proxy servers and the sale of services that can be used to game social media platforms, defraud ad networks, build botnets, and effectively launder illegal activity. We use the clustering from HMM to help identify which kinds of forums have high activity related to proxy servers. Compared to many of the other topics, topic 1.Locations %(as reported in Table~\ref{tab:interesting_topics_english})
is concerned primarily with the sale of proxy servers. When we examined cluster 2 from the HMM clustering, we noticed a significant activation of this topic. Interestingly, the ten most active forums in cluster 2 seem to capture most of the activation of the topic over time. As seen in Fig.~\ref{fig:proxy-topic}, we examine the forum-weeks (points) with significant probability (above 0.50) for topic 1.Locations from forums in cluster 2 and recover approximately 11,000 documents which seem to be automated posts advertising the sale of access to proxies all over the world. These 11,000 documents come mostly from CSU and BlackWorld: both forums have subforums dedicated to the advertising of proxies and are in the top 10 active forums in the cluster. An excerpt from one of these posts is as follows: \say{...camarillo | ca | unknown | united states | checked at vn5socks.netlive...}. When we look for documents pertaining to more specific uses of proxies, we find approximately 20 posts that directly mention ``viewbot''. Viewbotting is the act of using bots to artificially inflate the number of views on a social media profile (e.g. YouTube and Twitch). As it can be difficult to determine if a viewer is a human or a bot, this can potentially trick the social media platform into thinking a profile is more popular than it actually is and result in more attention than it would obtain organically. Alternate uses of viewbots are to watch video ads on a channel, artificially increasing ad revenue, etc.
An excerpt from one of these documents is as follows:
\say{...i viewbotted my vid to 1k views and got 10 slaves...}

Another potential malicious use of proxy servers is for carding. Carding in this case refers to the fradulent use of other people's credit cards, personal and/or financial data to purchase goods, launder money, or generally steal an individual's money. Proxy servers are commonly used to ``cash out'' stolen credit card information by buying items like pre-paid gift cards through payment processors. In our corpus, there are a number of documents that mention carding. An example that suggests intent to use proxies for carding is as follows:
\say{...proxies are often blacklisted when used for fraud so im looking for a source for fresh proxies for use for carding...}

\subsubsection{Marketplace Shutdowns}

% \begin{figure}[tbh]
%     \centering
%     \includegraphics[width=\columnwidth]{images/week_same_forum_cross_entropy_drugs.png}
%     \caption{Marketplace Shutdown Cross-entropy }
%     \label{fig:cross-entropy-markets}
% \end{figure}

%\begin{figure}
%    \centering
%    \includegraphics[width=\columnwidth]{images/weekly_entropy_drugs.png}
%    \caption{Weekly Entropy of Marketplace-related Forums}
%    \label{fig:entropy-drugs}
%\end{figure}

Our last case study regards the seizure of the AlphaBay marketplace by the FBI on July 4th 2017 and the seizure of the Hansa marketplace on July 20th 2017 by the Dutch NHTCU. We show cross-entropy, forum activity and transitions between states to analyze this case study.%demonstrate differences between the HMM results and the cross-entropy. 

\begin{figure*}%[bh]
    \centering
    \begin{tabular}{cc}
    \includegraphics[width=0.9\columnwidth]{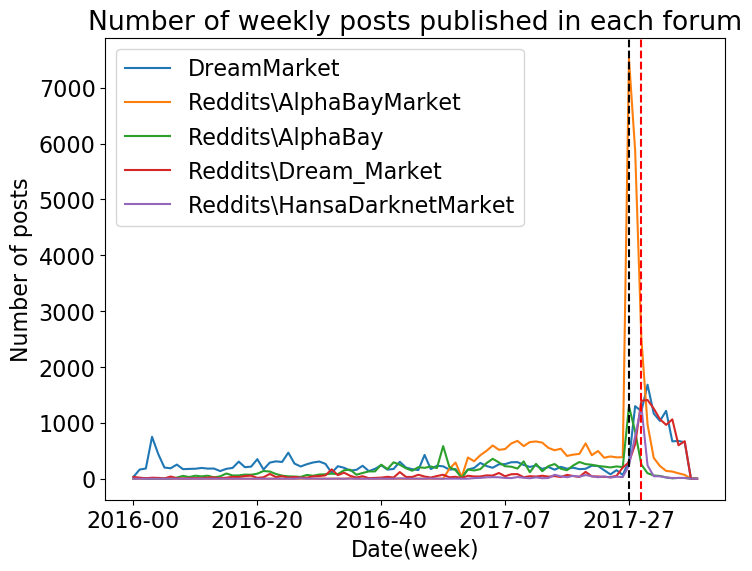}
    &   
    \includegraphics[width=0.9\columnwidth]{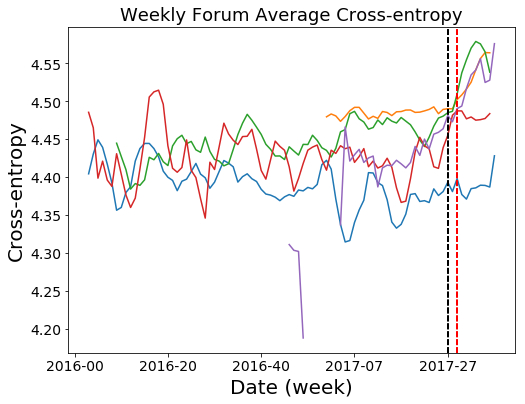} \\
    (a) & (b)
    \end{tabular}
    \caption{(a) Activity of forums relevant to AlphaBay and Hansa closure. (b) Smoothed weekly topic cross-entropy of forums. Cross-entropy is smoothed using a rolling average over 4 weeks. For both figures the black line indicates July 4th, 2017 when AlphaBay was seized. The red line indicates July 20th, 2017 when Hansa was seized.} 
    %\caption{Smoothed Weekly Topic Cross Entropy of Forums Forums affected by the AlphaBay and Hansa seizures in July 2017. The first black line indicates July 4th, 2017 when Alpha Bay was seized. The second red line indicates July 20th, 2017 when Hansa was seized. Smoothed using a rolling average over 4 weeks.}
    \label{fig:cross-entropy-most-active-bottom}
\end{figure*}

Included in our d2web data are forums related to transactions and reviews of this marketplace, including several private subreddits. We observed that a few of the forums in our dataset have peaks in activity around the same time. Figure~\ref{fig:cross-entropy-most-active-bottom}(a) shows number of posts published in these forums. %The dotted black line which is in line with when subreddits AlphaBay Market and AlphaBay had their highest value shows the date of the AlphaBay closure and the red dotted line in line with when subreddit Hansa Darknet Market had its highest value shows the date of the Hansa closure. 
Forums related to AlphaBay and Hansa have peaks on the date of AlphaBay and Hansa closure respectively. Interestingly, a week after the Hansa closure Dream Market and the subreddits DreamMarket and DreamMarketDarknet had their highest value which suggests that users of these two big market places, AlphaBay and Hansa, have migrated to Dream Market. To check whether the forums respond to these events by changing the topics of discussion, we compute the cross-entropy. %between the topic vector for the target forum for $week_{t}$ (averaged over all posts for that week) and the aggregate topic vector (averaged over all posts on the forum).
 As seen in Figure~\ref{fig:cross-entropy-most-active-bottom}(b) we see that two relevant forums, one about AlphaBay and one about DreamMarket, experience a change in their topic vectors after the shutdown (2017 week 27, or 07-04-2017).  The cross-entropy increases around that time, suggesting growing difference from the aggregate topic distribution for their respective forums however changes in the volume of posts were more significant. We also observe state transitions in Dream Market forum in all three dates (AlphaBay and Hansa closure and the week after when there is a peak in activity for Dream Market related forums). 
% \note{nathan: Do we need a sentence that says "we used this instead of HMM because XYZ"}

%\note{Nathan: Visualize topic change dynamics (Method 1 or 2) for the banned forums}

%\note{Nazgol: with HMMs can you discover forums where activity migrated post shutdowns }

%Interestingly, in Fig. ~\ref{fig:drug-topic}, we see numerous more forum-weeks that had high probability in the drug related topic. Looking at the forums represented in the set of most significant points, we see approximately 1000 documents from a forum devoted to DreamMarket, and approximately 60 documents from two AlphaBay related forums. Of note are 75 documents from forum 225, all of which seemed to be posts where vendors are introducing themselves in what seems to be a templated manner. 

%{Forums active on the topic of Rx. Do they also have illegal drug topics? What other topics do they have? Anything else?}
%\begin{figure}
%    \includegraphics[width=0.9\columnwidth]{images/shutdown.png}
%    \caption{Activity of forums relevant to AlphaBay and Hansa shutdown. The black and red vertical lines indicate AlphaBay and Hansa shutdown respectively.}
%    \label{fig:alphabay_activity}
%\end{figure}

%As we can see there is a wide variety of uses for proxy servers reflected in the dataset. Recently, there have been reports of the Ngioweb malware, a multifunctional proxy server, being used in botnet campaigns, making understanding the multiple different uses of proxies important \cite{Ramnit}. 

\section{Discussion \& Conclusions}

We used LDA to learn a rich set of latent components from a large corpus of documents spanning different topics and make use of a non-parametric HMM to better understand how those forums relate to one another in terms of the dynamics of their content. We then proceeded to use what we learn from both the states and the topics to identify specific posts discussing malicious activities that are understood to largely come from the dark and deep web. 

This work can be extended in a number of different ways. First, one can use our framework to analyze new and unseen forums. Second, we can extend the framework to explicitly consider patterns of activity (e.g., frequency of posts) alongside the semantic information of the topic dynamics. Additionally, using average of the topic distributions of posts in a forum in one week to represent that forum will invariably smooth out events that have a small number of events associated with them. An alternate method that would be more sensitive to smaller fluctuations is to use the maximum value for a topic obtained from any post in that week.

%It would be appropriate to compare the results of a system like ours against a Dynamic Topic Model which would allow for the topics themselves to change overtime.
A promising extension to our system would be to use a dynamic topic model that incorporates a birth-death process for the population of topics, allowing for new topics to emerge and for old topics to die off. Online communities move fast, and a topic model that includes data from even a couple months ago can become obsolete and hinder productive analysis. 

%To better understand the behavior going on in complex online communities, it is essential that we develop the tools to take a heterogeneous dataset (e.g, multiple different forums in multiple different languages) and extract relevant information. We have shown that we can learn a small number of relevant areas of interest in such a dataset, and then track how these areas of interest change over time, not to mention which forums are focused on which areas. %Identifying relevant areas of interest, and from that relevant data sources, 
%We can filter out irrelevant data sources from the dataset and increase our ability to identify relevant anomalous activity. 

\footnotesize{\textbf{Acknowledgments}. This work was supported by the Office of the Director of National Intelligence (ODNI) and the Intelligence Advanced Research Projects Activity (IARPA) via the Air Force Research Laboratory (AFRL) contract number FA8750-16-C- 0112, and the Defense Advanced Research Projects Agency (DARPA), contract number W911NF-17-C-0094. The U.S. Government is authorized to reproduce and distribute reprints for Governmental purposes notwithstanding any copyright annotation thereon. Disclaimer: The views and conclusions contained herein are those of the authors and should not be interpreted as necessarily representing the official policies or endorsements, either expressed or implied, of ODNI, IARPA, AFRL, DARPA, or the U.S. Government.
}

\bibliographystyle{ACM-Reference-Format}
\balance 
\bibliography{sample-base}

\end{document}